\documentclass{article}

\usepackage{docmute}
\usepackage{siunitx}
\usepackage{tabularx}

\usepackage{arxiv}

\usepackage[utf8]{inputenc}
\usepackage[T1]{fontenc}

\usepackage{amsmath}
\usepackage{amssymb}
\usepackage{amsfonts}
\usepackage{amsthm}
\usepackage{bm}
\usepackage{comment}
\usepackage{fancybox}
\usepackage{framed}
\usepackage{color}
\usepackage[dvipdfmx]{graphicx}
\usepackage{multicol}
\usepackage{multirow}
\usepackage{pdflscape}
\usepackage{verbatim}
\usepackage{subfigure}
\usepackage{cite}
\usepackage{hyperref}
\usepackage{url}

\hypersetup{
    colorlinks = true,
    urlcolor = blue,
    linkcolor = red,
    citecolor = green
}

\usepackage{algorithm}
\usepackage[noend]{algpseudocode}
\usepackage{algorithmicx}

\expandafter\def\expandafter\UrlBreaks\expandafter{\UrlBreaks
  \do\a\do\b\do\c\do\d\do\e\do\f\do\g\do\h\do\i\do\j%
  \do\k\do\l\do\m\do\n\do\o\do\p\do\q\do\r\do\s\do\t%
  \do\u\do\v\do\w\do\x\do\y\do\z\do\A\do\B\do\C\do\D%
  \do\E\do\F\do\G\do\H\do\I\do\J\do\K\do\L\do\M\do\N%
  \do\O\do\P\do\Q\do\R\do\S\do\T\do\U\do\V\do\W\do\X%
  \do\Y\do\Z}

\algrenewcommand\algorithmicindent{1.0em}

\algnewcommand\algorithmicforeach{\textbf{for each}}
\algdef{S}[FOR]{ForEach}[1]{\algorithmicforeach\ #1\ \algorithmicdo}
\algdef{S}[FOR]{ForEachParallel}[1]{\textbf{for each}\ #1\ \textbf{in parallel do}}

\algnewcommand\AlgAnd{\textbf{and} }
\algnewcommand\AlgOr{\textbf{or} }
\algnewcommand\AlgContinue{\textbf{Continue}}

\algrenewcommand\textproc{}

\algnewcommand{\Initialize}[1]{
	\State \textbf{Initialize:}
 	\State \hspace*{\algorithmicindent}\parbox[t]{0.8\linewidth}{\raggedright #1}}

\algnewcommand{\LeftComment}[1]{
    \Statex $\triangleright$ #1 \hfill}

\usepackage{breakurl}

\title{Performance Improvement of Federated Learning Server using Smart NIC}

\author{
  Naoki Shibahara\\
  Keio University\\
  3-14-1 Hiyoshi, Kohoku-ku, Yokohama, Japan\\
  \texttt{shibahara@arc.ics.keio.ac.jp}\\
  \And
  Michihiro Koibuchi\\
  National Institute of Informatics\\
  2-1-2 Hitotsubashi, Chiyoda-ku, Tokyo, Japan\\
  \texttt{koibuchi@nii.ac.jp} \\
  \And
  Hiroki Matsutani\\
  Keio University\\
  3-14-1 Hiyoshi, Kohoku-ku, Yokohama, Japan\\
  \texttt{matutani@arc.ics.keio.ac.jp} \\
}

\begin{document}

% Header
\maketitle

% Abstract and keywords

% abst.tex

\begin{abstract}
Federated learning is a distributed machine learning approach where local weight parameters trained by clients locally are aggregated as global parameters by a server.
The global parameters can be trained without uploading privacy-sensitive raw data owned by clients to the server.
The aggregation on the server is simply done by averaging the local weight parameters, so it is an I/O intensive task where a network processing accounts for a large portion compared to the computation.
The network processing workload further increases as the number of clients increases.
To mitigate the network processing workload, in this paper, the federated learning server is offloaded to NVIDIA BlueField-2 DPU which is a smart NIC (Network Interface Card) that has eight processing cores.
Dedicated processing cores are assigned by DPDK (Data Plane Development Kit) for receiving the local weight parameters and sending the global parameters.
The aggregation task is parallelized by exploiting multiple cores available on the DPU.
To further improve the performance, an approximated design that eliminates an exclusive access control between the computation threads is also implemented.
Evaluation results show that the proposed DPDK-based federated learning server on the DPU with the approximation accelerates the execution time by 1.39 times with a negligible accuracy loss compared with a baseline server on the host CPU.
\end{abstract}
%\begin{IEEEkeywords}
%Federated Learning, Smart NIC, DPU, DPDK
%\end{IEEEkeywords}

\keywords{Federated Learning \and Smart NIC \and DPU \and DPDK \and 25GbE}

% Body

% intro.tex

\section{Introduction} \label{sec:intro}
Due to the proliferation of smartphones and Internet-of-Things (IoT)
devices, the volume of data generated in our life is continuously
increasing, and Artificial Intelligence (AI) technologies to exploit
such data are rapidly evolving.
At the same time, uploading personal data to servers increases the
concern about data privacy.
To address this issue, federated learning \cite{Federated} is a
promising distributed machine learning approach that does not upload
privacy-sensitive raw data to servers.

In the federated learning, clients download a model from a server and
train it locally.
Then the trained weight parameters are sent to the server.
The server aggregates the local parameters and sends back the
aggregated global parameters to the clients.
The weight parameters are exchanged between the server and clients
multiple times during the federated learning.
The communication workload on the server increases as the number of
clients increases or the size of the model becomes larger.
Nevertheless, the aggregation process is not computationally heavy
since it is simply averaging the local parameters received from
clients.
Thus, it is an I/O intensive task with a high network processing workload
compared with the computation.

To mitigate the network processing workload, in this paper, the federated
learning server is offloaded to NVIDIA BlueField-2 DPU \cite{DPU}
which is a smart NIC (Network Interface Card) that has eight
processing cores.
Dedicated processing cores are assigned for receiving the local weight
parameters and sending the global parameters.
The server is implemented as user-space application with DPDK (Data Plane
Development Kit) \cite{dpdk} so that a network protocol stack of Linux
kernel is bypassed, resulting in a lower processing latency and
higher network throughput.
The aggregation task is parallelized by exploiting multiple cores
available on the DPU.
To further improve the performance, an approximated design that
eliminates an exclusive access control between the computation threads
is implemented.
The baseline server and the approximated server are evaluated in terms
of the execution time and learning convergence to show the performance
and accuracy tradeoffs.

The rest of this paper is organized as follows.
Section \ref{sec:related} introduces background knowledge about the
federated learning, DPDK, and smart NIC.
Sections \ref{sec:design} and \ref{sec:implementation} describe the
design and implementation of the proposed federated learning server on
the smart NIC, respectively.
Section \ref{sec:eval} evaluates it in terms of the execution time and
learning convergence.
Section \ref{sec:conc} summarizes this paper and mentions our future
work.

% related.tex

\section{Background and Related Work} \label{sec:related}
\subsection{Federated Learning}
\begin{figure}[t]
    \centering
    \includegraphics[height=55mm]{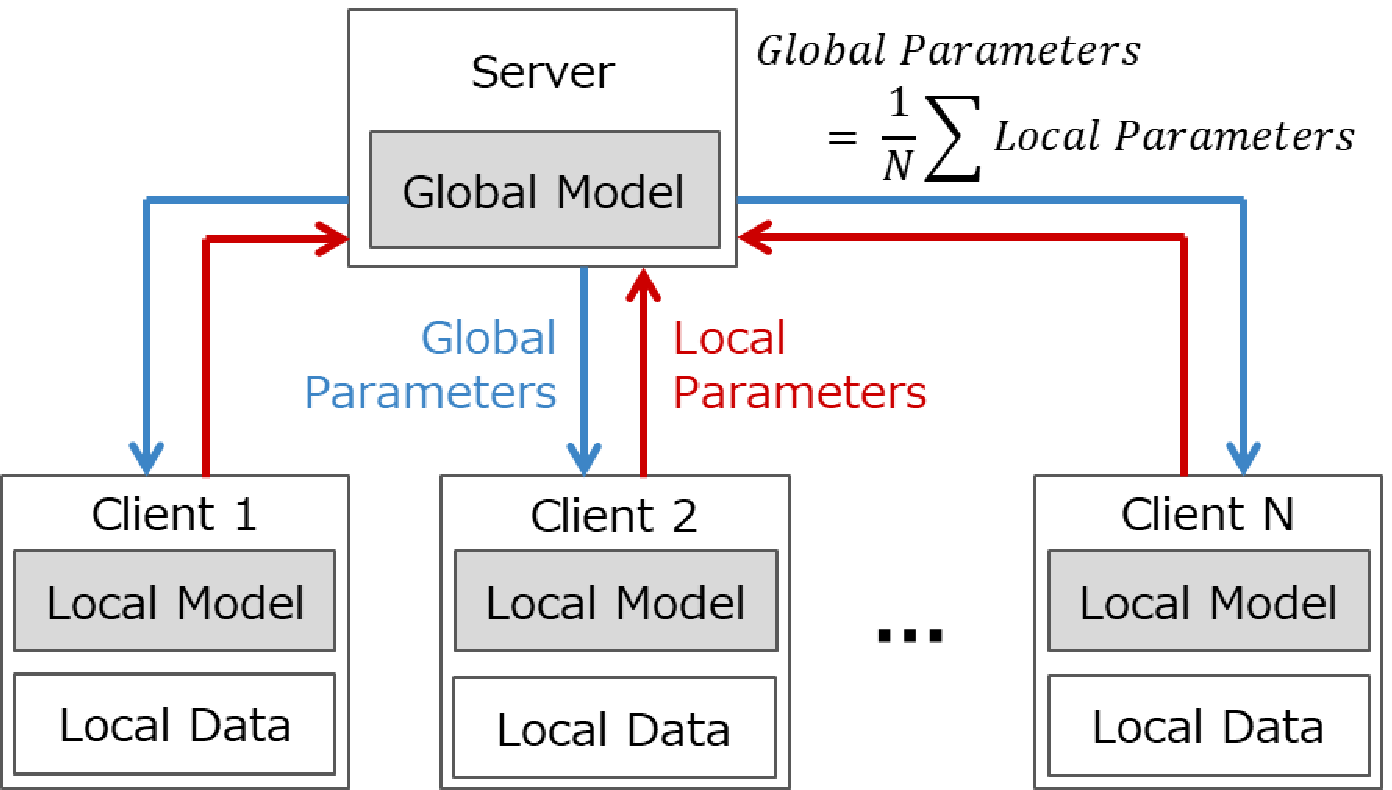}
    \caption{Basic federated learning system}
    \label{fig:fl}
\end{figure}

Modern mobile devices such as smartphones are major sources of
valuable data that can enhance user experiences while such personal
data are privacy-sensitive.
To obtain a global model trained from such privacy-sensitive data
owned by clients without uploading them to the server, the federated
learning \cite{Federated} has been extensively studied.
Figure \ref{fig:fl} illustrates a basic federated learning system with a
single server and $N$ clients.
Each client trains its local model using its own data and sends the
trained local parameters to the server.
The server aggregates the local parameters to produce global
parameters, which are then sent back to the clients.
Thus, the clients can share their trained results by incorporating
the global parameters in their local parameters.

\subsubsection{Federated Averaging}
FedAvg (Federated Averaging) is a typical federated learning
algorithm, and it is shown in Algorithm \ref{alg:federated-averaging}.
First, weight parameters of the target model are initialized.
For each round, $m$ clients are randomly selected, where $C$ is a
probability that a client is selected.
The selected clients join the aggregation, while the others keep
their local weight parameters.
In round $t$, a selected client $k$ trains its local parameters
$w^k_t$ using its own local data, based on the formula in line
\ref{alg:update}, to produce $w^k_{t+1}$.
The trained local parameters $w^k_{t+1}$ are sent to the server.
The server averages the received local parameters, based on the
formula in line \ref{alg:average}, to produce the aggregated global
parameters $w_{t+1}$.
In the algorithm, $n$ is the total number of data samples, and $n_k$
is the number of data samples owned by client $k$.
The global parameters $w_{t+1}$ are sent back to the clients, and each
client updates its local parameters using the received global
parameters.
These steps are repeated for $T$ rounds to obtain the final global and
local parameters.

\begin{algorithm}[h]
  \caption{Federated Averaging \cite{Federated}.
    $K$ clients are indexed by $k$.
    $B$ is local minibatch size, $E$ is number of local epochs, and
    $\eta$ is learning rate.}
    \label{alg:federated-averaging}
    \begin{algorithmic}[1]
        \Function{ExecuteServer}{\null}
            \State Initialize $w_0$
            \ForEach{round $t = 1, 2, \ldots$}
                \State $m \gets \max(C \cdot K, 1)$
                \State $S_t \gets \text{(random set of $m$ clients)}$
                \ForEachParallel{client $k \in S_t$}
                % \ForEach{client $k \in S_t$}
                    \State $w_{t + 1}^k \gets$ \Call{ClientUpdate}{$k, w_t$}
                \EndFor
                \State $w_{t + 1} \gets \displaystyle \sum_{k = 1}^K \
                \frac{n_k}{n} w_{t + 1}^k$ \label{alg:average}
            \EndFor
        \EndFunction

        \Function{ClientUpdate}{$k, w$} \hfill $\triangleright$ \textit{Run on client $k$}
            \State $\mathcal{B} \gets \text{(split $\mathcal{P}_k$ into batches of size $B$)}$
            \ForEach{local epoch $i$ from $1$ to $E$}
                \ForEach{batch $b \in \mathcal{B}$}
                    \State $w \gets w - \eta \nabla \ell(w; b)$ \label{alg:update}
                \EndFor
            \EndFor
        \EndFunction
    \end{algorithmic}
\end{algorithm}

\subsubsection{Advanced Federated Learning Algorithms}
In FedAvg algorithm, clients replace their local parameters with
the global parameters entirely every round.
It is effective when the goal is to improve the global model accuracy
for an entire data distribution that all clients will encounter in
future.
In reality, however, data distribution may differ depending on client
environments.
For such non-i.i.d. (independently and identically distributed) data,
Per-FedAvg \cite{perfed} and APFL \cite{apfl} are representative
algorithms to improve the local model accuracy for the local data
distribution of each client.
Although these algorithms have a similarity to FedAvg algorithm at the
point that the global parameters are obtained by averaging local
parameters, clients update their local parameters through a
weighted average of both the local and global parameters.
Please note that the server-side averaging process of FedAvg can also
be used in these advanced algorithms.

\subsection{DPDK}
DPDK provides libraries and network drivers to accelerate packet
processing.
More specifically, network processing is executed as a user-space
application that bypasses the network protocol stack of OS kernel.
Dedicated CPU cores are assigned for receiving packets, on which the
user-space applications are polling the NIC to receive packets.
This can reduce the overheads for context switching and data copying
compared with a network protocol stack of OS kernel triggered by
interrupts, resulting in a lower processing latency and higher network
throughput.
Please note that utilization of CPU cores that are polling the NIC is
always 100 percent.
% 2023/07/03 Matsutani added
In addition, the memory access can be accelerated by utilizing
hugepages supported by Linux.
Their sizes are larger than standard $4\si{\kilo\byte}$ pages so that
TLB (Translation Lookaside Buffer) misses can be reduced.
Since they are statically allocated in a physical memory, page-in and
page-out overheads can also be eliminated.

% 2023/07/03 Matsutani added
There are some high-performance network processing frameworks that
support TCP/IP on top of DPDK.
F-Stack \cite{fstack}, DPDK-ANS \cite{ans}, and mTCP \cite{mtcp} are
open-source network processing frameworks based on TCP/IP stack of
FreeBSD on top of DPDK.
These frameworks assume the shared nothing architecture, in which data
are not shared between multiple processing cores.
In this case, incoming packets are distributed to each processing core
by Receive Side Scaling (RSS) of NICs and then the packets are
processed within the assigned core.
ZygOS \cite{zygos} and Shenango \cite{shenango} support a shared
memory so that multiple processing cores can share data.
A task scheduler that can distribute the workload to multiple cores
to balance their workload is also implemented.
In this paper, the federated learning server workload is predictable
since the number of clients and the model size are determined
beforehand.
We can thus balance their workload by assigning the same number of
clients for each processing core.

\subsection{Smart NIC}\label{ssec:smartnic}
Smart NIC is a kind of NICs that have processing cores to perform
custom packet processing and routing control functions.
These tasks are typically executed by the CPU of the host machine.
By offloading them to the smart NIC, the CPU workload of the host machine
is reduced, so that the host CPU can concentrate on the other user
applications.
% In recent years, networking technologies are developing rapidly.
For example, VPNs (Virtual Private Networks) employ packet
encapsulation and encryption to ensure a secure data
communication.
In this case, adding encapsulation headers increases the packet sizes,
and the encryption and decryption increases the computation
overheads.
In addition to the standard packet processing and routing functions,
intrusion detection from external networks \cite{Wu2022}, 
data encryption/decryption, and data compression/decompression can 
be offloaded to the smart NICs.
As the networking technology continues to evolve, it becomes
increasingly complex.
Smart NICs have a good potential to offload such network processing to
the NIC and reduce the CPU workload of the host machine.

% 2023/07/03 Matsutani added
In \cite{offloading_dist}, various smart NIC products are evaluated in
detail.
Generally, processing cores implemented on the smart NICs are slower
than those of host CPUs.
Also, their L2/L3 caches and DRAMs are not abundant.
% This means that smart NICs may not be an ideal platform for offloading
% tasks that impose a high MPKI (L2 Cache Misses Per Kilo Instructions)
% or use a large memory.
In \cite{offloading_dist}, application characteristics that can be 
efficiently offloaded to smart NICs are analyzed, and a task scheduling 
that considers this insight is proposed.
In this paper, we focus on the aggregation process of the federated
learning server, in which a shared memory that can store only a single
set of global parameters is used and its memory access pattern is
straightforward.
This suggests that using smart NICs is a promising solution to offload
the federated learning server.

In this paper, we use NVIDIA BlueField-2 DPU MBF2H332A-AENOT as a
target smart NIC.
It is comprised of an SoC (System-on-Chip) that includes an 8-core ARM
processor running at 2.5GHz, 16GB DRAM, two 25Gbit Ethernet (GbE)
interfaces, and PCIe Gen4 interface.
It is connected to the network via the 25GbE interfaces and connected
to the host machine via the PCIe Gen4 interface.
Operating systems such as Linux is running on the DPU, and DPDK
applications can be executed on it.

In \cite{offload_dnn}, a data augmentation task is offloaded to the DPU in
order to accelerate deep learning tasks by overlapping the data
augmentation performed on the DPU and other training steps performed
on the host CPU.
Since the data augmentation does not require a network processing and
the DPU is used as an additional computation resource, benefits of the
25GbE interfaces of the DPU are not fully exploited.
In this paper, on the other hand, the federated learning server is
offloaded to the DPU as an I/O intensive task, so it can fully utilize
the benefits of smart NICs.

% 2023/07/03 Matsutani added
In \cite{novel} and \cite{characterizing}, a host CPU workload is
offloaded to the DPU by exploiting the RDMA (Remote Direct Memory Access)
functionalities.
Specifically, in \cite{novel}, communication primitives that support
non-blocking point-to-point communication and collective communication
are implemented on the DPU.
In \cite{characterizing}, communication performance between DPU and
host CPU via PCIe and that of RDMA are studied.
As a case study, KVS (Key-Value Store) is implemented on DPU.
Specifically, a part of keys in the KVS is cached in local DRAM of the
DPU to accelerate the KVS application.
Please note that the federated learning server running on the DPU in
this paper is quite simple.
It processes incoming packets directly and returns the aggregated
results to clients without communicating with the host CPU via PCIe.

% 2023/10/15 shibahara added
% \textcolor{red}{
% In this paper, we offload the aggregation process of federated
% learning to DPU.
% It is effective to reduce the CPU workload of the host machine.
% Since OS works in DPU, it is easy to change the algorithm.
% Additionally, we propose a communication protocol using DPDK and an
% approximate proces for speedup, and achive the aggregation process
% that is 1.39 times faster than the process by the host CPU.
%}
In this paper, we offload the aggregation process of federated
learning onto the DPU.
It is effective to reduce the CPU workload of the host machine. 
Other than DPU, using FPGA (Field Programmable Gate Array) based
programmable NICs is another solution to offload the CPU workload.
However, since Linux and development tools/libraries are available on
the DPU, the portability of software programs is high, which is
attractive especially for machine learning tasks.

% design.tex

\section{Federated Learning Server on DPU} \label{sec:design}

\begin{figure}[t]
    \centering
    \includegraphics[height=45mm]{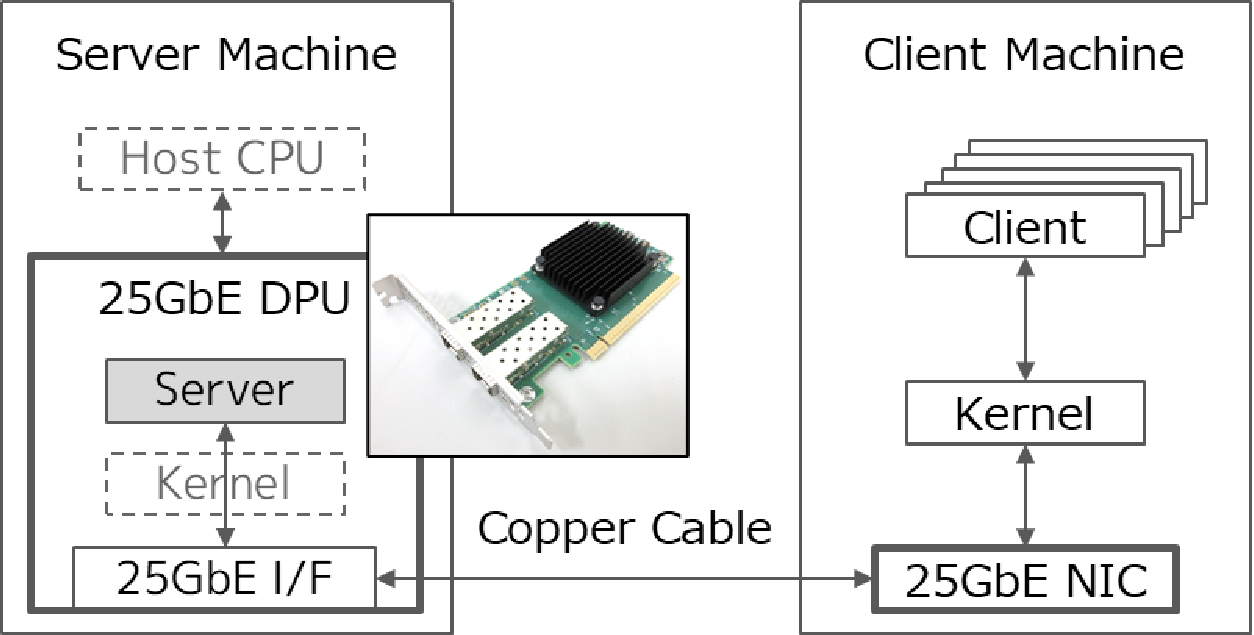}
    \caption{Federated learning system where federated learning server
      is running on DPU}
    \label{fig:fl_system}
\end{figure}

Figure \ref{fig:fl_system} illustrates a federated learning system
consisting of a single server and clients.
In this paper, the server is running on the DPU.
They are connected by a copper cable.

\subsection{Client Process}
Each client has its local model and trains it with its own local data.
The local training is repeated several times, and then the client
sends the trained weight parameters to the server via the 25GbE
network.
After the client receives the aggregated global parameters from the
server, it updates its local parameters by substituting them with the
received global parameters.
These steps are referred to as a ``round'' in federated learning.
By repeating these rounds, we can lower the training error for the
local data and share the local training results to the other
clients.

To exchange the weight parameters between the server and clients, we
use UDP as a lightweight transport layer protocol as will be described
more specifically in Section \ref{sec:ack}.
It is compared with a baseline implementation that uses TCP.
In both the cases, the client processes are implemented in Python.
They use socket APIs for the communication with TCP or UDP via a
TCP/IP protocol suite of the OS kernel.
In the case of packet loss, the missing global parameters are
complemented with the local parameters.
In other words, the missing part is left as the local parameters.

\subsection{Server Process}
The federated learning server receives local parameters from 
clients and aggregates them based on Algorithm
\ref{alg:federated-averaging}.
That is, the local parameters are averaged to produce global
parameters.
The global parameters are then sent back to the clients.
The federated learning server is implemented in C++.
The baseline server uses socket APIs for the TCP communication via
a TCP/IP protocol stack of the OS kernel.
The proposed server running on the DPU uses UDP communication, and it
is implemented with DPDK.
It directly accesses the NIC and handles the UDP communication without
using the protocol stack of the OS kernel.
It thus parses the Ethernet, IP, and UDP headers of incoming packets
and generates outgoing packets in the DPU.

\subsubsection{DPDK Model}
DPDK applications can be modeled as ``Run to Completion'' model or
``Pipeline'' model.
In the Run to Completion model, packet reception, packet processing,
and packet transmission steps are executed by a single logical CPU
core.
On the other hand, they are partitioned and executed by multiple
logical cores in the Pipeline model.
In the Run to Completion model, the packet processing step becomes a
bottleneck if the processing is complicated and time-consuming,
resulting in a lower overall throughput.
In this paper, we employ the Pipeline model that distributes the
processing steps to multiple processing cores in order to eliminate 
the performance bottleneck.

\subsubsection{Packet Processing} \label{sec:flow}
\begin{figure}[t]
    \centering
    \includegraphics[height=50mm]{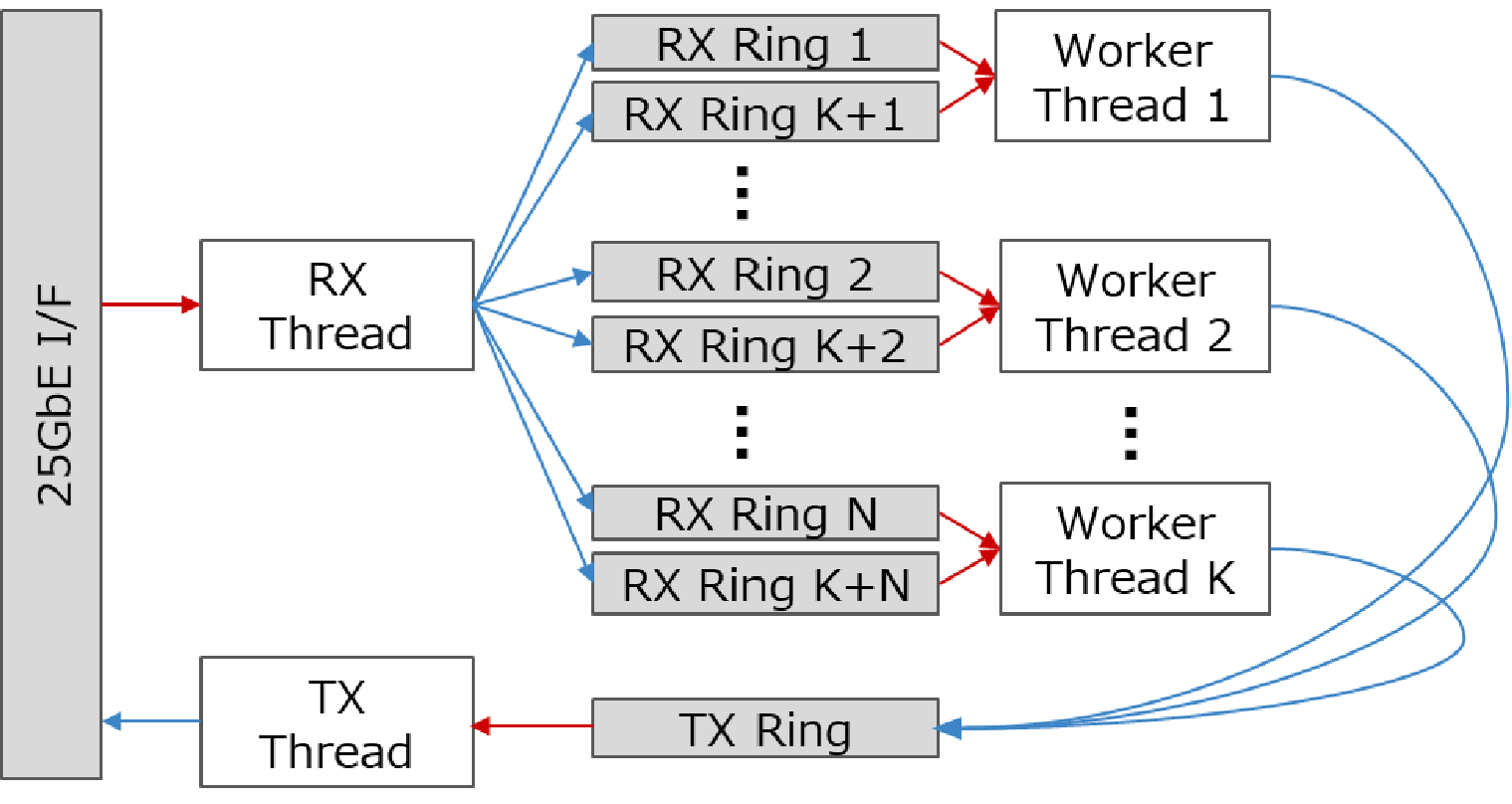}
    \caption{Multiple threads of server process on DPU}
    \label{fig:pipeline}
\end{figure}

Figure \ref{fig:pipeline} illustrates the server process based on the
Pipeline model.
An RX thread, $K$ Worker threads, and a TX thread are connected via
ring queues.
Each thread is executed on a dedicated processing core.
Red arrows indicate dequeueing of packets, while blue ones are
enqueueing of packets.
These ring queues employ the Ring library provided by DPDK for
communication between the threads.
They enqueue and dequeue pointers of mbuf objects, which represent
packets, to exchange packets between these threads.
The RX thread polls the packet receiving queue of the NIC.
When a packet is received, it verifies the Ethernet and IP headers of
the packet to confirm that it is the packet coming from a federated
learning client.
It also checks the source port number in the UDP header and puts the
packet into the RX ring of the corresponding client.
Additionally, the RX thread sends acknowledgment packets to clients as
will be described in Section \ref{sec:ack}.

Worker thread $i$ sequentially polls RX ring queues $i, i+K, i+2K, ...$.
Once a packet is retrieved from the RX ring queue, local parameters
are extracted from its payload and added to a float array which has
been initialized with 0.
After local parameters of all the clients have been added to the float
array, a single Worker thread divides each element of the float array by the
number of clients in order to calculate element-wise averages of the
local parameters (i.e., new global parameters).
Local parameters that are missing due to packet loss are not included
in the divisor.
In other words, the number of clients participating in the aggregation
may differ depending on the element of the global parameters.
The other Worker threads wait until the element-wise averages are
calculated using a spinlock mechanism.

After the averages are calculated, each Worker thread copies the
global parameters to payload of packets for clients, fills out
Ethernet, IP, and UDP headers of the packets, and puts them into the
TX ring.
The TX thread polls the TX ring.
Once a packet is retrieved from the TX ring, it is enqueued to the
transmission port of the NIC and sent to the client.
The TX thread then releases the mbuf objects, in which the packet was
stored.

% and returns them to the mempool.
% In the DPU platform, multiple processing cores are available.
% The aggregation task is parallelized by exploiting multiple cores
% available on the DPU.
% Dedicated threads are assigned for receiving the local weight
% parameters and sending the global parameters.
% To further improve the performance, an approximated server that
% eliminates an exclusive access control between the computation
% threads is also implemented.

\subsubsection{Lightweight Network Protocol} \label{sec:ack}
In DPDK, since a network protocol stack of OS kernel is bypassed, a
flow control functionality has to be provided by the application
layer.
A client process has two communication states: sending local
parameters and receiving global parameters.
Similarly, a server process has three states: receiving local
parameters, computation, and sending global parameters.
To guarantee the correct state transitions, a simple yet reliable
network protocol that uses acknowledgement packets is implemented in
the application layer.
For example, to detect the completion of local parameter reception, a
server may be able to count the total number of packets received.
However, if a packet is lost, the server may still be in the state of
receiving local parameters while the client has already finished
sending local parameters and has transitioned to the state of
receiving global parameters.
In this case, the federated learning may stop.

To avoid such a situation, the clients send an acknowledgement packet
denoted as ``END'' to indicate the end of transmission after sending
the local parameters.
The server then responds to the clients with a response packet denoted
as ``END ACK''.
The client keeps sending END packets until an END ACK packet is
received, so that it can ensure that the server has finished receiving
the local parameters.
The client transitions to the state of receiving global parameters
after the reception of END ACK.
In our implementation, two types of control packets, namely START and
END, and their corresponding response packets, namely START ACK and
END ACK are used.

\begin{figure}[t]
    \centering
    \includegraphics[height=75mm]{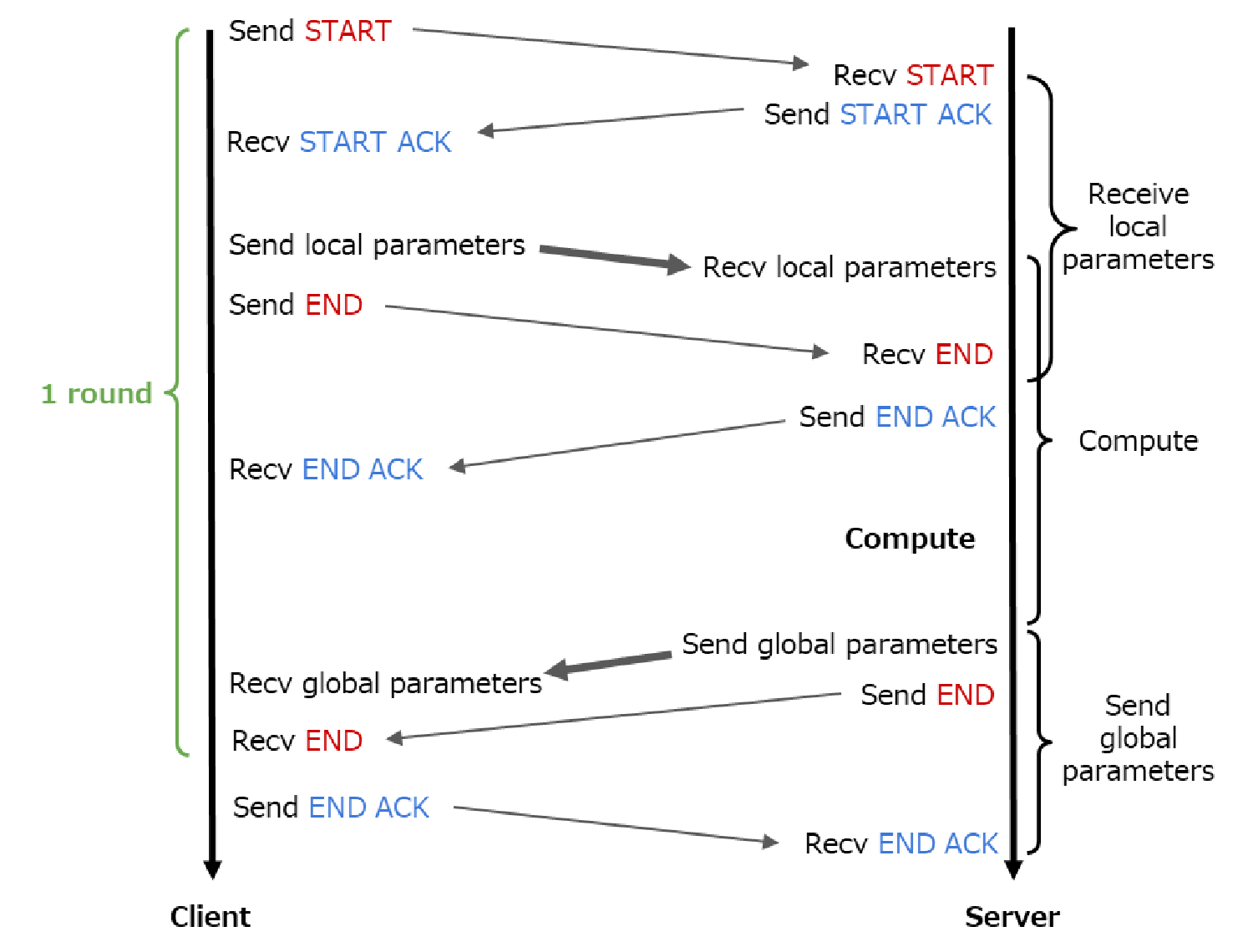}
    \caption{Lightweight protocol for UDP communication}
    \label{fig:control}
\end{figure}

Figure \ref{fig:control} illustrates the communication protocol
between a client and a server.
When the client completes its local training, it sends a START packet
to the server.
Upon receiving a START ACK packet from the server, the client sends
local parameters and then an END packet.
When an END ACK packet is received from the server, the client
terminates the state of sending local parameters and transitions to
the state of receiving global parameters.
After receiving the global parameters, when an END packet is received
from the server, the client responds with an END ACK packet.
The client then replaces the local parameters with the received global
parameters, and completes a single round of federated learning.
% The reception of END packet and the transmission of END ACK packet are
% expected during a time period when the local parameters are replaced.
Since there is a possibility that the END ACK packet is lost,
retransmitted END packets can be handled for one second after the first END packet is received.
TCP also has a similar waiting period when terminating a connection.
RFC 793 defines this period as twice the MSL (Maximum Segment
Lifetime), which is commonly two minutes.
In our communication protocol, the waiting period is set to this minimum 
value, though this did not affect the evaluation results in
this paper.

When the server receives a START packet, it responds with a START ACK
packet and then receives local parameters until an END packet is
received from the client.
As described in Section \ref{sec:flow}, the RX thread enqueues
incoming packets to the RX rings immediately after receiving them.
Worker threads poll the RX rings to retrieve the packets as soon
as they arrive, and then the element-wise addition of the received
local parameters is executed.
That is, the reception and addition of local parameters are performed
in parallel until an END packet is received.
When an END packet is received, the server waits until the Worker
threads process all the packets stored in the RX rings and then
terminates the state of receiving local parameters.
The server then transitions to the state of computation by responding
with an END ACK packet, so that it performs element-wise division to
the float array to obtain the global parameters.
When the RX thread receives an END packet, it puts an END ACK packet
directly into the TX ring without passing the packet to the Worker
thread.
If the END ACK packet is lost at this point, the RX thread can receive
another END packet retransmitted by the client while the server has
transitioned to the state of computation.
This implementation can reduce the number of context switches between
these threads.
After the element-wise division is completed, the server sends the
global parameters to each client.
The global parameters are sent in the same way as the local
parameters.
A single round is then completed once an END ACK packet is received
from the clients.

\subsubsection{Elimination of Exclusive Access Control}
In this server design, multiple Worker threads running on multiple processing
cores execute the element-wise addition on the same float array stored
in the shared memory of the DPU.
There is a possibility of write-write conflicts (e.g., lost update)
between multiple Worker threads.
To avoid the conflicts, an exclusive access control mechanism is
typically required for the threads to ensure precise computation
results.
To further improve the performance, in this paper an approximated
federated learning server that eliminates this exclusive access
control is also implemented.
In Section \ref{sec:eval}, the baseline server and the approximated
server are evaluated in terms of the execution time and federated
learning convergence to show the performance and accuracy tradeoffs
between them.

% implementation.tex

\section{Implementation Details} \label{sec:implementation}
\subsection{Client Process} \label{sec:impl_client}
% In this paper, we use FedTorch \cite{FedTorch} as a federated
% learning platform.
% We built FedTorch with PyTorch 1.14.0a0 and Open MPI 4.1.4.
% In FedTorch, as many client processes as the number of clients train
% their local model, and they perform inter-process communication to
% aggregate the models using MPI (Message Passing Interface).
% In our implementation, we disable the inter-process communication and
% use socket APIs instead.
% Each client process sends local parameters to the server by using
% socket APIs from Python.

% In the baseline implementation, a network protocol stack of OS kernel
% is used for the communication.
% As a transport layer protocol, the proposed approach uses UDP, while 
% it is compared with a baseline TCP version in Section \ref{sec:eval}.
% In both the cases, the client processes use socket APIs from Python.

\begin{figure}[t]
    \centering
    \includegraphics[height=35mm]{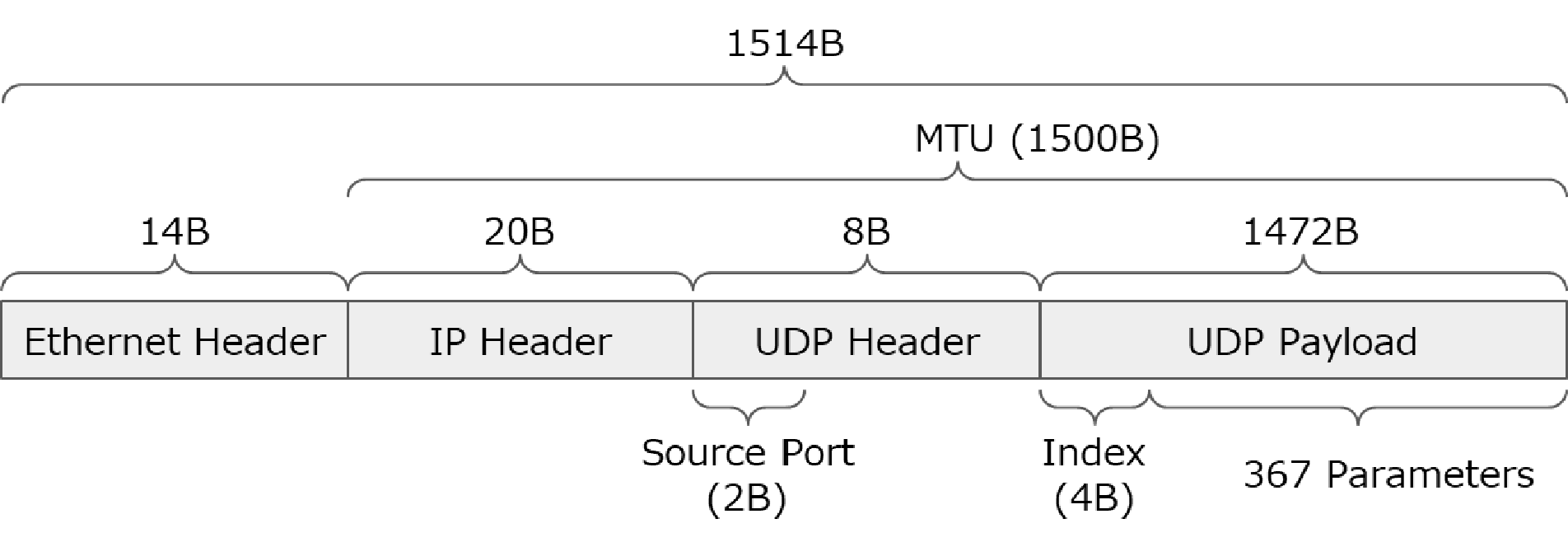}
    \caption{Packet format for UDP communication}
    \label{fig:header_udp}
\end{figure}

Figure \ref{fig:header_udp} illustrates the packet format for the UDP
communication.
Each client is identified by source port number of the UDP header.
In the client process, the local parameters are extracted from the
trained local model and converted to numpy.float32.
Then, they are serialized and transmitted to the server.
To detect packet loss and guarantee the ordered data transfer, a
$4\si{\byte}$ index number of packets is added to the beginning of
each payload.
The payload size without the index number is $1468\si{\byte}$ if we
assume MTU (Maximum Transmission Unit) is $1500\si{\byte}$ and IP and
UDP headers are $20\si{\byte}$ and $8\si{\byte}$, respectively.
In this case, each UDP packet can convey $367$ weight parameters.

\subsection{Server Process}
As mentioned in Section \ref{ssec:smartnic}, we use NVIDIA BlueField-2
DPU MBF2H332A-AENOT as a platform of the proposed DPDK-based federated
learning server.
In this section, the baseline and proposed servers are described
below.
% \subsubsection{DPU Specification}
% We use NVIDIA BlueField-2 DPU MBF2H332A-AENOT.
% As mentioned in Section \ref{ssec:smartnic}, it is comprised of an SoC
% that contains an $8$-core ARM processor running at
% $2.5\si{\giga\hertz}$, $16\si{\giga\byte}$ DRAM, two $25\si{\giga bE}$
% interfaces, and PCIe Gen4 interface.
% It is connected to the network via $25\si{\giga bE}$ interfaces and
% connected to the host machine via PCIe Gen4 interface.

\subsubsection{Baseline TCP Server}
%ホストCPUでの実装はクライアント数分のプロセスが作られ、コンテキストス
%イッチすること
Here, we describe a baseline implementation of federated learning
server using TCP.
It uses socket APIs provided by OS kernel for communication.
It is compared with the proposed DPDK server using UDP in Section
\ref{sec:eval}.
In the TCP communication, MSS (Maximum Segment Size) is set to
$1460\si{\byte}$ to comply with MTU of the UDP communication, which is
$1500\si{\byte}$.
The overall packet length including the Ethernet header is
$1514\si{\byte}$ in both cases.

The baseline TCP implementation runs either on the host CPU or the
$8$-core processor of DPU.
When a TCP connection request is received from a client, a new thread
is created through the use of the standard library std::thread.
The assignment of threads to cores is done by the OS kernel.
Each thread receives local parameters from the client and subsequently
performs the element-wise addition of the local parameters to the
float array.
Only a single thread executes the element-wise division to produce
global parameters, while the other threads wait for the division by
using std::mutex and std::condition\_variable, which are provided by
the standard library.

% TSO (TCP Segmentation Offload) is a function that allows the NIC to
% carry out TCP segmentation instead of the kernel.
% The kernel can transmit and receive packets larger than the MTU,
% thus reducing the load on the kernel.
% GSO (Generic Segmentation Offload) is a software-based
% implementation of TSO that performs TCP segmentation when TSO is not
% supported by the NIC.
% In this study, these functions are disabled since the packet length
% is manually determined.

\subsubsection{DPDK Server}
In the proposed DPDK server implementation, the global parameters are
declared as a conventional float array.
By employing the operator+= of std::atomic\_ref$<$float$>$, which is
an atomic reference to a float variable implemented in C++20, an
exclusive access control is enforced only during the execution of the
element-wise addition.
On the other hand, since the division operation is carried out by a
single representative Worker thread, it is executed without using
std::atomic\_ref$<$float$>$.
As mentioned in Section \ref{sec:design}, if the exclusive access
control between these threads is eliminated, the precise averages 
cannot be guaranteed.
We implement such an approximated server without using
std::atomic\_ref$<$float$>$.
It is compared with the baseline DPDK server that uses the exclusive
access control in terms of the speed and learning convergence.

% eval.tex

\section{Evaluations} \label{sec:eval}
\subsection{Evaluation Environment}
Table \ref{tab:env} shows the evaluation environment of server and
client machines.
The DPU is attached to the server machine via PCIe Gen4 interface as a
NIC.
The DPU and client machine are connected by a 2m 25GbE direct attach
copper cable.
The server process is executed either on the DPU or the server
machine.
When the server process is executed on the DPU, packets coming from
the client machine to the DPU's physical interface are forwarded to
the ARM processor of the DPU.
In this case, federated learning packets are not forwarded to the
server machine (i.e., host CPU) since the aggregation process is
entirely offloaded onto the DPU.
On the other hand, when the server process is executed on the host
CPU, the packets are forwarded from the DPU's physical interface to
the host machine's interface.

In this experiment, the dataset is CIFAR-10, which consists of 50,000
training samples and 10,000 test samples.
The number of client processes is ten.
The training samples are partitioned to the ten client processes
equally, so each client process has 5,000 i.i.d. training samples.
The global model is tested with the 10,000 test samples.
The model architecture is CNN consisting of four convolutional layers
and two fully connected layers
\footnote{Conv(32, 3) $\to$ Relu $\to$ Conv(64, 3) $\to$ Relu $\to$
Maxpool(2) $\to$ Conv(128, 3) $\to$ Relu $\to$ Conv(256, 3) $\to$
Relu $\to$ Maxpool(2) $\to$ FC(256) $\to$ Dropout(0.5) $\to$ FC(10)
$\to$ Softmax(10)}.
The number of parameters is about two million, and each parameter is
represented as a 32-bit float.

The server process is executed on the 8-core ARM processor of the
DPU, in which two cores are dedicated to the RX and TX threads,
respectively.
Worker threads are executed on five cores.
Since there are ten clients, each thread handles two clients.
One core is left for other tasks including the OS task scheduling.
The client processes are implemented with Python 3.11.4, Pytorch 
2.0.1, and torchvision 0.15.2.
The server process is implemented with C++ and DPDK 20.11.3, and 
compiled with -O3 optimization level.

\begin{table}[t]
	\centering
	\caption{Specification of server machine, DPU, and client machine}
	\begin{tabular}{llll}
	  \hline
	   & Server Machine & DPU & Client Machine \\
	  \hline
	  OS & Ubuntu 20.04 & Ubuntu 20.04 & Ubuntu 20.04 \\
	  CPU & Intel Core i7-11700 & ARM Cortex-A72 & Intel Core i7-10700 \\
	  %    & (8C) & (8C) & (8C) \\
	  RAM & 16 GB & 16 GB & 32 GB \\
	  % Pytorch/Open MPI & $-$ & $-$ & 1.14.0a0 / 4.1.4 \\
	  NIC	     & NVIDIA BF-2 DPU & $-$ & Intel XXV710-DA2 \\
	  % DPDK     & 21.11.2 & 20.11.3 & $-$ \\
	  \hline
	\end{tabular}
	\label{tab:env}
\end{table}

\begin{figure}[t]
    \centering
    \includegraphics[height=70mm]{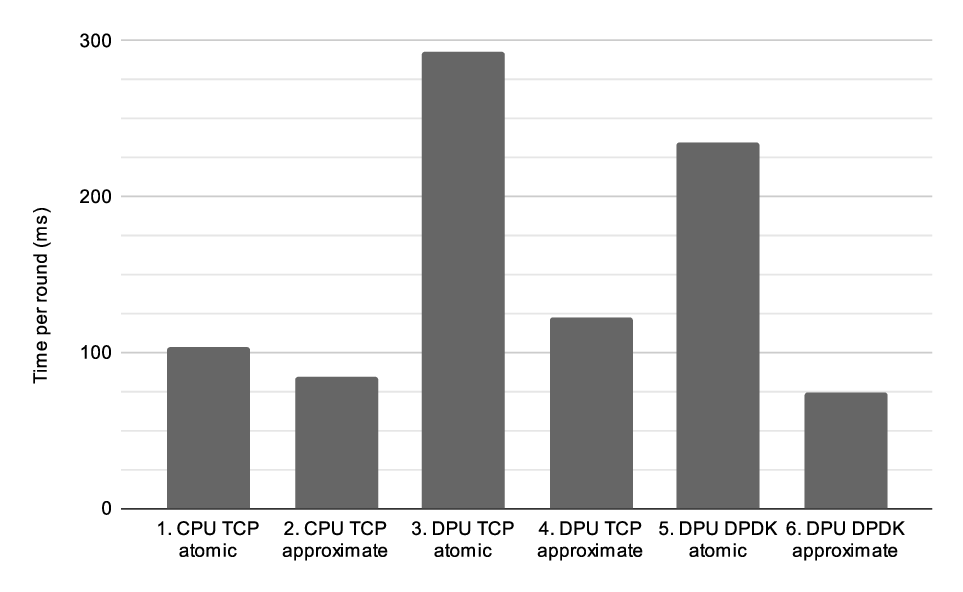}
    \caption{Server response time measured in client}
    \label{fig:eval_client}
\end{figure}
\begin{figure}[t]
    \centering
    \includegraphics[height=70mm]{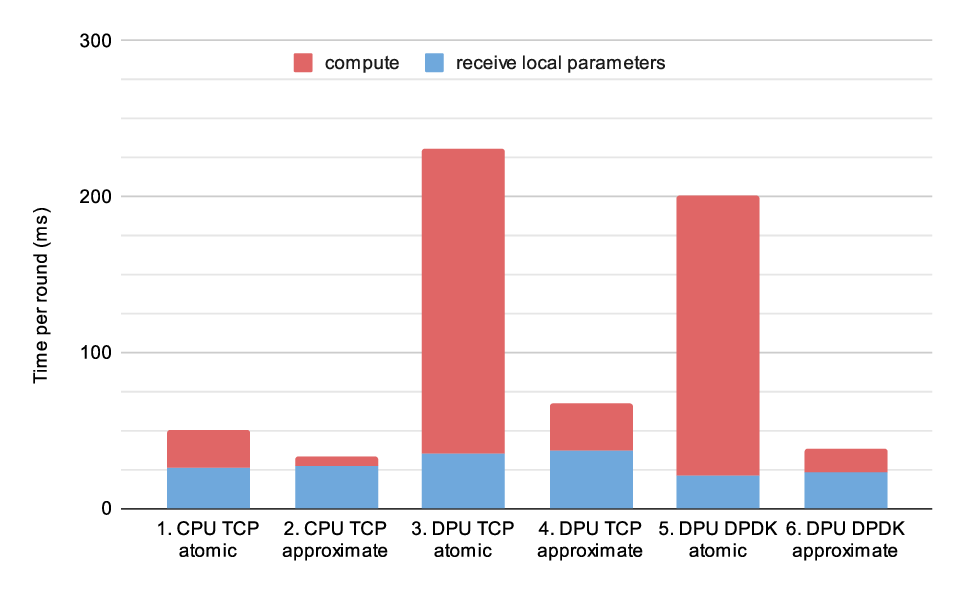}
    \caption{Server execution time measured in server}
    \label{fig:eval_server}
\end{figure}

\begin{figure}[t]
    \centering
    \includegraphics[height=80mm]{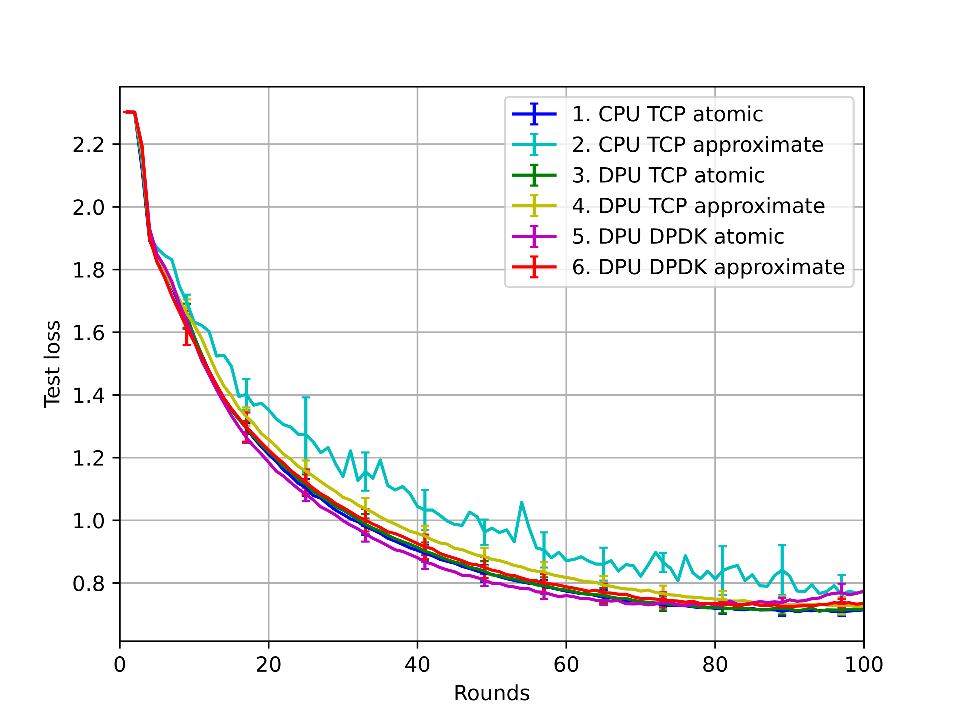}
    \caption{Training convergence (rounds vs. test loss)}
    \label{fig:loss}
\end{figure}

\subsection{Evaluation Results}
As shown in Figure \ref{fig:control}, a client sends a START packet to
a server, and then it receives an END packet from the server.
The proposed federated learning server is evaluated in terms of the
latency to receive the END packet after the START packet is sent.
This is the server's response time for the aggregation which is
observed by the client.

Figure \ref{fig:eval_client} shows the evaluation results of the
following six server implementations.
\begin{enumerate}
\item Server running on host CPU using TCP/IP protocol stack with
  exclusive access control
\item Server running on host CPU using TCP/IP protocol stack without
  exclusive access control
\item Server running on DPU using TCP/IP protocol stack with exclusive
  access control
\item Server running on DPU using TCP/IP protocol stack without
  exclusive access control
\item Server running on DPU using DPDK with exclusive access control
\item Server running on DPU using DPDK without exclusive access control
\end{enumerate}

In addition, these servers are evaluated in terms of the latency to
complete the parameter aggregation after a START packet is received.
Figure \ref{fig:eval_server} shows the evaluation results.
The blue bar shows the receiving time of local parameters, which means
the latency to receive the END packet after the START packet is
received.
Then, the element-wise addition of received local parameters and the
element-wise division of accumulated local parameters are executed.
The red bar shows the computation time for the addition and division.
Please note that Figure \ref{fig:eval_server} shows the latencies
measured at the server side; so the transmission time of the global
parameters is not included.
The complete latencies including the transmission time are shown in
Figure \ref{fig:eval_client}.

Regarding (1) and (3), although they are the same program, the program
execution on the DPU is much slower than that on the CPU.
Especially, the computation time (red part) is increased in the DPU
since processor performance of DPU is lower than that of the host CPU.
Regarding (3) and (4), their difference is the exclusive access
control.
Eliminating the exclusive access control speeds up the computation
time of global parameters by 6.66 times.
The comparison between (1) and (2) also shows a similar tendency while
the speedup is smaller than that on the DPU.
Regarding (3) and (5), their difference is the implementation of the
communication; (3) uses a standard TCP/IP stack while (5) uses the
proposed DPDK-based optimized communication.
Using the DPDK-based optimized communication, the receiving time of
local parameters at server is improved by 1.65 times, and the
aggregation response time is improved by 1.25 times from the client's
view.
The server computation time for global parameters is also slightly 
improved (i.e., 1.09 times speed up).
This is because a part of computation (red part) is overlapped with
the parameter reception (blue part) as mentioned in Section
\ref{sec:ack} and thus reduced.
% 2023/07/03 Matsutani added
Utilizing hugepages may also contribute the performance improvement.
The proposed approach (6) combines the elimination of exclusive access
control and use of the DPDK-based optimized communication implemented
on the DPU.
The proposed approach improves the aggregation response time by 3.93
times compared with (3).
It also improves the response time by 1.39 times compared with (1)
which is executed on the host CPU.

\subsection{Training Convergence}
Figure \ref{fig:loss} shows training curves of the six
approaches at a client.
We conducted the same experiments five times.
The X-axis represents the rounds, while the Y-axis represents the
average and standard deviation of the test loss.
Although the approximated computation on the baseline CPU
implementation introduces fluctuations in the training curve as shown
in (2), the negative impact of the approximation is small in (4) and
negligible in (6).
Since the performance and parallelism of the host CPU are higher than
those of DPU, it is expected that (2) introduces more
write-write conflicts and fluctuations.
Although the DPDK-based UDP communication introduces packet loss
especially in the global parameter transfer from the server to clients
(e.g., 4.68\% in (6)), since our lightweight protocol can handle the
packet loss, the accuracy loss is limited.
As a result, the training curve of the proposed approach (6) is very
close to the CPU baseline (1).

% conc.tex

\section{Summary} \label{sec:conc}
Although the federated learning algorithm is an emerging research
topic and continuously becoming sophisticated, the major computation
task at the server side is typically averaging the received local
parameters.
In this I/O intensive task, the network processing accounts for a
large portion compared to the computation.
In this paper, we implemented the aggregation process of the federated
learning server on NVIDIA BlueField-2 DPU as a smart NIC.
Although offloading the federated learning server on the DPU can
mitigate the host CPU workload, a simple offloading increases the
execution time compared with that on the host CPU due to its lower
processor performance.
Our approach thus combines the elimination of exclusive access control
and use of the DPDK-based lightweight communication implemented on the
DPU.
The experiment results showed that the proposed approach significantly
improves the aggregation response time compared with the DPU baseline
and it is even higher than the host CPU baseline by 1.39 times.
Training curve of the proposed approach using CIFAR-10 dataset on the
DPU showed a similar learning convergence to the CPU baseline.
Further investigations on more practical environments using non
i.i.d. datasets are our future work.

%本論文の欠点を挙げるのは重要ですが、ちょっと書きすぎかな。。
%本研究では10 個のクライアントが1 台のクライアントマシン上で動作するため、連合学習システム全体のネットワーク処理性能は1 台のクライアントマシンのネットワーク処理性能に律速されてしまう。
%連合学習は本来1 台のデバイスが1 つのクライアントであり、多数のクライアントマシンに対して1 台のサーバマシンが存在する。
%物理的なクライアントマシンを増やし、通信のボトルネックをサーバにすることで、本研究で実装した連合学習サーバのスループットを測定できるだろう。
%クライアント数の増加により通信量が増大し、より頻繁にパケットロスが発生するようになった場合の学習の収束への影響を調べる必要がある。
%Additionally, employing existing DPDK-based TCP protocol stacks, such as F-Stack, may result in more optimized communication.

{\bf Acknowledgements} 
This paper is based on results obtained from ``Research and Development Project of the Enhanced Infrastructures for Post 5G Information and Communication Systems'' JPNP20017)), commissioned by the New Energy and Industrial
Technology Development Organization (NEDO).

% Bibliography
%\bibliographystyle{unsrt}
%\bibliography{refer}

\end{document}